\documentclass[aps,pre,onecolumn]{revtex4}
\usepackage{amsfonts,amssymb,amsmath,bm}
\usepackage[dvips]{graphicx}
\usepackage{color}
\usepackage[T1,T2A]{fontenc}
\usepackage[cp1251]{inputenc}
\usepackage[english,russian]{babel}

\begin{document}

\title{Scale dependent chirality in twist-bend liquid crystals.}

\author{E.I.Kats}
\affiliation {Laue-Langevin Institute, Grenoble, France (Retired), and \\
Landau Institute for Theoretical Physics, RAS, \\ 
Chernogolovka, Moscow region, Russia (Retired)}


\begin{abstract}

The discovery of new classes of lyotropic and thermotropic liquid crystals (e.g., twist-bend, splay, and ferroelectric phases), together with significant advances in experimental techniques for their investigation, has renewed interest in a number of physical phenomena that have been studied in classical liquid crystals (nematics, cholesterics, and smectics) for more than a century.
In this paper, we revisit one such ''old--new'' problem, recently highlighted in the preprint by A. Ashkinazi, H. Chhabra, A. El Moumane, M. M. C. Tortora, and J. P. K. Doye, ''Chirality Transfer in Lyotropic Twist-Bend Nematics,'' arXiv:2508.03544v1 (2025), in which various mechanisms of chirality transfer from the molecular scale to the structural scale were discussed.
Here we present a simple theoretical analysis of chirality transfer within a Landau theory describing the phase transition between cholesteric ($Ch$) and chiral twist-bend ($N_{TB}^*$) liquid crystals. We demonstrate that the handedness of the heliconical $N_{TB}^*$ structure is opposite to that of the parent cholesteric phase. This relationship originates from the orthogonality between the cholesteric director and the vector order parameter characterizing the $N_{TB}^*$ phase.

\end{abstract}


\maketitle

{\bf {Preface.}}${\, }$
This work is dedicated to the memory of the outstanding theoretical physicist Philippe Nozi\`eres, with whom I had the privilege of discussing numerous scientific and broader intellectual questions during my time (2000--2012) in the Theory Group at the Institut Laue--Langevin (ILL).
For purely chronological reasons, I never had the opportunity to discuss the present manuscript with Philippe. Nevertheless, during our almost daily conversations, I came to appreciate what I believe was the defining feature of his approach to theoretical physics. Philippe viewed theoretical physics as a unified science describing real materials and the physical phenomena occurring within them. He believed that the same universal theoretical principles could—and should—be applied to systems ranging from low-temperature metals to liquid crystals at room temperature.

This paper is offered as a small tribute to his memory and as an expression of my gratitude.

{\bf {Introduction.}}${\, }$

The relatively recently discovered class of liquid crystals with a tilted nematic-like structure, commonly known as twist-bend nematics, continues to attract considerable attention owing to its rich fundamental physics and promising potential applications \cite{HS09,PN10,CD11,BK13,MD14}. Particularly remarkable is the fact that this phase exhibits a heliconical orientational order, and is therefore chiral, despite being formed from achiral molecules.
Note to the point, to illustrate Ph. Nozieres view of theoretical physics as a unified science 
that a closely related sequence of symmetry breaking occurs in superfluid $He^{3}$ confined in nematically ordered aerogel, where the normal state first undergoes a transition to a polar, nematic-like phase and subsequently to the polar-distorted A phase. The latter transition introduces an additional orbital component orthogonal to the pre-existing anisotropy axis and  produces a chiral order parameter \cite{MD19}.

By comparison, conventional nematic liquid crystals ($N$) possess uniaxial, achiral orientational order, whereas cholesteric liquid crystals, known for more than a century, exhibit a simple orthogonal helical structure with a pitch typically of several micrometers. The cholesteric structure results from weak molecular chirality, which explains its relatively large pitch, while the handedness of the helix is determined by the sign of the molecular chirality. In contrast, the $N_{TB}$ phase emerges through spontaneous chiral symmetry breaking and is characterized by a nanoscale pitch.

Recently, Ref.~\cite{AC25} reported the discovery of the $N_{TB}$ phase in lyotropic liquid crystals. Unlike the thermotropic case, the phase transition is athermal, being controlled by the volume fraction of mesogenic (''banana-shaped'') molecules dissolved in an isotropic solvent rather than by temperature. Moreover, the lyotropic nature of these systems makes it possible to investigate mixtures of chiral and achiral mesogens. In Ref.~\cite{AC25}, a chiral twist-bend phase, $N_{TB}^*$, was observed and identified.

In this case, chirality is already broken in the parent nematic-like phase, namely the cholesteric ($Ch$) phase. Consequently, the two possible heliconical states of the $N_{TB}^*$ phase (clockwise and counterclockwise) are no longer equally probable at the $Ch \rightarrow N_{TB}^*$ phase transition. An especially intriguing experimental observation reported in Ref.~\cite{AC25} is that the handedness of the $N_{TB}^*$ heliconical structure is opposite to that of the parent cholesteric phase from which it develops.

The discovery of the lyotropic $N_{TB}^*$ phase naturally raises the question of whether analogous thermotropic $N_{TB}^*$ liquid crystals may also exist. In this paper, we formulate a Landau-type theory describing the $Ch \rightarrow N_{TB}^*$ phase transition. Our model generalizes the Landau theory previously developed in Ref.~\cite{KL14} for the transition from an achiral nematic ($N$) phase to the twist-bend ($N_{TB}$) phase by incorporating molecular chirality. We demonstrate that the theory predicts opposite handedness for the cholesteric and chiral twist-bend helices, in complete agreement with the experimental observations reported in Ref.~\cite{AC25}.

A conceptually related problem involving the interplay between two competing types of chirality on different length scales was investigated in Refs.~\cite{KA96}--\cite{KL00}. The considerably more sophisticated systems studied in those works include chiral hexatic columnar phases formed by long polymer molecules. In those examples, the two types of chirality are frustrated and become compatible only through the nucleation of topological defects. By contrast, for the $Ch \rightarrow N_{TB}^*$ phase transition considered here, the chiralities of the two structures are compatible without the need for topological defects.

{\bf {Landau Theory of the $Ch$--$N_{TB}^*$ Phase Transition.}}${\, }$

We begin by considering a liquid crystal in the cholesteric ($Ch$) phase, where the director
\[
{\bf n}\rightarrow
n_x=\sin(q_0z),\qquad
n_y=\cos(q_0z),\qquad
n_z=0,
\]
provides the soft long-wavelength degrees of freedom.
In contrast, the modulation of the $N_{TB}^*$ phase is short-ranged, occurring on the molecular length scale $a$, with the characteristic wave vector satisfying
$q_s a\simeq 1$,
whereas the cholesteric modulation is characterized by the much longer wavelength condition
$q_0a\ll 1$.

Since the modulation in the $N_{TB}^*$ phase is short-ranged, the corresponding order parameter is also a short-wavelength quantity.
However, this short-wavelength order parameter develops on the background of the long-wavelength orientational order already present in the cholesteric phase, which is described by the unit director ${\bf n}$.
Consequently, the free-energy expansion should contain the following contributions \cite{KL14,GP93,KL06}:

\begin{itemize}

\item
The conventional long-wavelength Frank free energy,

\begin{eqnarray}
 {\cal F}_\mathrm{Fr}=\int dV\
 \left\{ \frac{K_1}{2} (\nabla\cdot\bm n)^2
 +\frac{K_2}{2}\left[{\bm n}\cdot curl {\bm n}\right ]^2
 +K_2q_0\left[{\bm n}\cdot curl {\bm n}\right ]
 +\frac{K_3}{2} (\partial_z \bm n)^2 \right\}.
 \label{Frank}
 \end{eqnarray}

\item

By definition, the short-wavelength component of the order parameter,
$\bm\varphi$, must be orthogonal to the long-wavelength director,
\[
\bm n\cdot\bm\varphi=0.
\]
Thus, the vector $\bm\varphi$ has two independent components and serves as the order parameter of the $Ch$--$N_{TB}^*$ phase transition.

The corresponding Landau functional cannot contain terms that are odd in $\bm\varphi$, since $\bm\varphi$ is a vector order parameter.
Therefore, within mean-field theory, the $Ch$--$N_{TB}^*$ transition is expected to be continuous (second order), and the Frank free energy (\ref{Frank}) must be supplemented by its short-wavelength counterpart.

\item

The local (non-gradient) contribution to the Landau free energy has the same universal form as for an ordinary long-wavelength order parameter,

\begin{eqnarray}
F_0 =
 \int dV \left\{\frac{a}{2} \bm\varphi^2
 +\frac{\lambda}{24} {\bm \varphi}^4
 \right\},
 \label{lr1}
 \end{eqnarray}

where, as noted above, odd powers of ${\bm\varphi}$ are forbidden by symmetry.

\item

In contrast, the gradient terms for the short-wavelength order parameter differ substantially from those describing a conventional long-wavelength order parameter.

We assume that the modulation wave vector of the $N_{TB}^*$ phase is oriented perpendicular to the local cholesteric director.
Then the leading contribution describing the softening of the order parameter near the two wave vectors
${\bf q}=\pm q_s\hat{\bf e}_z$
(where $\hat{\bf e}_z$ is the unit vector along the $z$ axis, chosen parallel to the unperturbed cholesteric axis) is

\begin{eqnarray}
 \int dV \left\{\frac{b_{||}}{8 q_{s}^2}
 \left[
 \left(n_i n_k \partial_i \partial_k + q_{s}^2\right)
 \bm\varphi
 \right]^2
 \right\}.
 \label{lr2}
 \end{eqnarray}

Because the gradients of the short-wavelength order parameter ${\bm\varphi}$ are not small, the contribution (\ref{lr2}) must be supplemented by the second-order gradient terms

\begin{eqnarray}
 \int dV \left\{
 \frac{b_1}{2}(\nabla\bm\varphi)^2
 +\frac{b_2}{2}\delta^\perp_{ij}
 \partial_i\bm\varphi
 \partial_j\bm\varphi
 \right\},
 \label{lr3}
 \end{eqnarray}

where
\[
\delta^\perp_{ij}=\delta_{ij}-n_in_j.
\]

The terms (\ref{lr3}) are of the same order as the quadratic contribution proportional to
$a{\bm\varphi}^2$.

Similarly, the fourth-order gradient contribution

\begin{eqnarray}
 \int dV \left\{
\frac{\lambda_1}{16 q_{s}^2}
 \left(
 \epsilon_{ijk}
 \varphi_i
 \partial_j
 \varphi_k
 \right)^2
 \right\},
 \label{lr4}
 \end{eqnarray}

is generally of the same order as the local quartic term
$(\lambda/24){\bm\varphi}^4$.

\end{itemize}

All of the above contributions (\ref{lr1})--(\ref{lr4}) describing the short-wavelength order parameter are achiral.
To describe the $Ch$--$N_{TB}^*$ phase transition, they must be supplemented by the chiral contribution to the short-wavelength free energy,

\begin{eqnarray}
 {\cal F}_\mathrm{Chiral}
 =
 \int dV
 \left\{
 \frac{\lambda_1}{16q_s}
 \varepsilon_{ijk}
 \varphi_i
 \nabla_j
 \varphi_k
 \right\}.
 \label{Chiral}
\end{eqnarray}

Having established the free-energy functional, one should, in principle, minimize the total free energy with respect to all admissible variations of both the cholesteric director ${\bm n}$ and the vector order parameter ${\bm\varphi}$ describing the $N_{TB}^*$ phase.

However, the objective of the present work is more modest: namely, to determine the relationship between the handedness of the cholesteric director field ${\bf n}$ and that of the vector order parameter ${\bm\varphi}$.
For this purpose, it is unnecessary to solve the complete system of Euler--Lagrange equations explicitly.
Instead, in the next section we shall use only the constraint relating ${\bm n}$ and ${\bm\varphi}$ that follows from these Euler--Lagrange equations.

{\bf {Chirality inheritance from $Ch$ into $N_{TB}^*$ phase.}}${\, }$

A direct, although somewhat cumbersome, minimization of the free energy presented above leads to the following conclusion.
At the phase transition from the cholesteric phase to the chiral $N_{TB}^*$ phase, heliconical domains with clockwise and counterclockwise twists are no longer energetically degenerate. This is in sharp contrast to the $N$--$N_{TB}$ phase transition, where the two domains have identical free energies, as expected for spontaneous chiral symmetry breaking.

More specifically, for a cholesteric helix with clockwise twist ($q_0>0$), the energetically preferred $N_{TB}^*$ domain is twisted in the opposite direction.
Likewise, for a cholesteric helix with counterclockwise twist ($q_0<0$), the lowest-energy $N_{TB}^*$ domain is twisted clockwise.

Rather than obscuring the underlying physics with lengthy algebra, the same result can be obtained by a simple argument that determines the handedness of the vector order parameter ${\bm\varphi}$ characterizing the chiral $N_{TB}^*$ phase.
{\color{red}{
The short-wavelength order parameter reads as
$$
\boldsymbol{\varphi}(z)
=
\varphi_0
\left\{
\cos(q_s z),\,
-\sin(q_s z),\,
0
\right\}.
$$

This form is explicitly orthogonal to the cholesteric director

$$
\mathbf n(z)=
\left\{
\sin(q_0z),\,
\cos(q_0z),\,
0
\right\},
$$

when the corresponding phase relation is imposed. In particular,

$$
\mathbf n\cdot\boldsymbol{\varphi}
=
\varphi_0\sin[(q_0-q_s)z],
$$

which leads after the minimization of of the free energy with respect to the admissible variations of $\bm\varphi$ 
subject to the constraint $|\bm n|=1$ directly to the condition:

\begin{eqnarray}
{\bm n}\cdot{\bm\varphi}=0,
\label{ch1}
\end{eqnarray}

i.e., the vector order parameter remains everywhere perpendicular to the cholesteric director.}}

For the simple cholesteric helix

$$
{\bf n}
=
\left\{
n_x=\sin(q_0z),\,
n_y=\cos(q_0z),\,
n_z=0
\right\},
$$

this orthogonality condition yields

\begin{eqnarray}
\sin(q_0z-q_sz)=0,
\label{ch2}
\end{eqnarray}

which is satisfied provided that

\begin{eqnarray}
(q_s-q_0) z =\pi N,
\label{ch3}
\end{eqnarray}

where $N$ is an arbitrary integer.

Let us first consider the case $q_0>0$, corresponding to a right-handed (clockwise) cholesteric helix.
Equation (\ref{ch3}) then implies that, for every point along the cholesteric axis ($z$ axis),

\begin{eqnarray}
q_sz>\pi N.
\label{ch4}
\end{eqnarray}

This relation shows that, for even values of $N$, the vector ${\bm\varphi}$ lies in the fourth quadrant of the $(x,y)$ plane, whereas for odd values of $N$ it lies in the third quadrant.
In both cases, however, the rotation of ${\bm\varphi}$ is counterclockwise.
Thus, the heliconical twist of the $N_{TB}^*$ phase has the opposite handedness to that of the parent cholesteric phase.

The opposite case is entirely analogous.
For a counterclockwise cholesteric helix ($q_0<0$), the vector order parameter ${\bm\varphi}$ rotates clockwise, corresponding to $q_s>0$.
Therefore, regardless of the sign of the cholesteric twist, the heliconical modulation in the $N_{TB}^*$ phase always possesses the opposite handedness.

Finally, we note that, for weakly chiral molecular building blocks, the macroscopic handedness of the cholesteric phase is expected to be opposite to the handedness of the constituent particles \cite{ST76}. In contrast, for strongly chiral molecules, the handedness of the macroscopic phase coincides with that of the molecular building blocks \cite{ST76}.

{\bf {Kapitza pendulum analogy for chirality inheritance.}}${\, }$

The Kapitza pendulum is a classical example of the dynamical stabilization of an inverted rigid planar pendulum whose pivot undergoes high-frequency vertical oscillations.
This phenomenon was first discovered by Stephenson \cite{ST08} and later independently rediscovered and analyzed by Kapitza \cite{KA51}.

Stabilization of the inverted pendulum occurs when the natural frequency of the pendulum, $\omega_0$, is much smaller than the frequency of the external oscillations, $\omega_s$, namely when
\[
\omega_0 \ll \omega_s.
\]

In chiral $N_{TB}$ liquid crystals, one encounters two very different spatial scales characterized by the wave vectors $q_0$ and $q_s$, with
\[
q_0 \ll q_s.
\]
Their ratio therefore serves as a small parameter that allows the construction of a phenomenological theory of chirality inheritance.

Although the Kapitza pendulum is a dynamical system, whereas chiral $N_{TB}$ liquid crystals are governed by equilibrium thermodynamics, the underlying physics and the mathematical descriptions of the two phenomena are remarkably similar. In both cases, the essential procedure consists of averaging over rapid oscillations: temporal oscillations for the pendulum and spatial oscillations for the twist-bend order parameter.

Furthermore, in both systems a critical condition exists at which the topology of the effective energy changes. For liquid crystals, this corresponds to a change in the structure of the free-energy landscape, whereas for the mechanical pendulum it corresponds to a change in the effective potential derived from the Lagrangian. In both cases, the transition occurs between a single-minimum and a double-minimum energy landscape.

In this brief note we do not present the full details of the averaging procedure over the rapid spatial oscillations of the $N_{TB}$ order parameter associated with the wave vector $q_s$. Instead, we only outline the logic of the approach, which closely parallels that used in the theory of the Kapitza pendulum.

After averaging over the short-scale modulation, one obtains the effective free energy

\begin{eqnarray}
F_{eff}=K_{eff}q_0^2+V_{eff}(\delta),
\label{kp1}
\end{eqnarray}

where $K_{eff}$ is the effective cholesteric Frank elastic constant renormalized by its coupling to the $N_{TB}$ order parameter ${\bm\varphi}$, and $\delta$ denotes the relative phase shift between the two helices: the orthogonal cholesteric helix and the tilted $N_{TB}$ helix.

A simple phenomenological form of the effective potential is

\begin{eqnarray}
V_{eff}(\delta)
=
-\alpha \cos(\delta)
+\beta \cos(2\delta),
\label{kp2}
\end{eqnarray}

where the coefficients $\alpha$ and $\beta$ are phenomenological parameters expressed in terms of the microscopic quantities describing the $Ch$--$N_{TB}$ phase transition:

\begin{equation}
(\alpha,\beta)
=
(\alpha,\beta)
\left(
\frac{q_0}{q_s},
A,
g,
K
\right),
\label{kp3}
\end{equation}

where $A$ is the amplitude of the twist-bend modulation, $g$ characterizes the coupling between the cholesteric director ${\bf n}$ and the vector order parameter ${\bm\varphi}$, and $K$ denotes the relevant elastic constant.

From Eq.~(\ref{kp2}) it is straightforward to see that, for
\[
\beta<\alpha/4,
\]
the effective free energy possesses a single minimum at
\[
\delta=0,
\]
corresponding to a co-rotating configuration of the cholesteric and $N_{TB}$ helices.

In contrast, for
\[
\beta>\alpha/4,
\]
the effective free energy develops two minima,
\[
\delta=0
\qquad \mathrm{and} \qquad
\delta=\pi.
\]
The second minimum corresponds to a counter-rotating configuration of the two helices.

The critical condition for this topological transition may be estimated in terms of the original parameters as

\begin{eqnarray}
g^2A^2\simeq Kq_s^2.
\label{kp4}
\end{eqnarray}

{\bf {Conclusion.}}${\, }$

To place the present work in the context of recent developments, we again refer to Ref.~\cite{AC25}, in which the authors investigated how chirality introduced through a twist in the center of the molecular building block is transferred to the macroscopic heliconical twist of the chiral $N_{TB}^*$phase.

Motivated by their results, we have investigated a different aspect of chirality transfer (hereafter, instead of the term ''transfer'' proposed in \cite{AC25}, we will use the term "inheritance'').  Namely the inheritance of chirality from the long-wavelength cholesteric director field to the heliconical order of the $N_{TB}^*$ phase. Our analysis shows why the handedness of the chiral heliconical $N_{TB}^*$ structure is opposite to that of the parent cholesteric phase.

At first sight, this result appears counterintuitive. Indeed, one might naively expect that the only distinction between the heliconical structure of the $N_{TB}^*$ phase and the parent cholesteric phase is the finite tilt of the director with respect to the helix axis. In both phases, the director ${\bf n}$ rotates as one moves along the helical axis. If chirality were determined solely by the sense of this rotation, one would naturally conclude that the handedness should remain unchanged. Although the director in the $N_{TB}^*$ phase possesses an additional bend component, the local structure remains helicoidal. The bend merely tilts the director away from the helix axis without reversing the direction of its rotation.

As is often the case in physics, however, this intuitive picture is incomplete. The transition from the cholesteric phase to the $N_{TB}^*$ phase is not simply a tilt of the long-wavelength director ${\bf n}$. Instead, it is driven by the condensation of the short-wavelength vector order parameter ${\bm\varphi}$ at the finite wave vector $q_s$. Consequently, the preferred handedness of the $N_{TB}^*$ phase is determined by the chiral coefficient $\lambda_1$, rather than by the direction in which the cholesteric director precesses.

Chirality on different length scales is therefore governed not only by purely geometric considerations, such as the sense of rotation of the cholesteric helix (pure twist) or the combined twist and bend deformations of the heliconical $N_{TB}^*$ structure. Because the order parameter of the $N_{TB}^*$ phase is a short-wavelength quantity, its chirality is determined primarily by molecular packing on nanometer length scales. In this respect, the situation is closely analogous to the relationship between molecular chirality and the handedness of the macroscopic helix in conventional cholesteric liquid crystals.

For ordinary cholesterics composed of weakly chiral molecules, molecular packing favors a macroscopic helix whose handedness is opposite to that of the constituent molecules \cite{ST76}. In contrast, the short-wavelength order parameter of the $N_{TB}^*$ phase corresponds to the regime associated with strongly chiral molecular building blocks, in which the handedness of the phase coincides with that of the local molecular packing \cite{ST76}.

At the level of a simple heuristic picture, one may say that the molecular packing responsible for the formation of the short-pitch heliconical structure, described by the order parameter ${\bm\varphi}$, energetically favors a reversal of the handedness with respect to the parent cholesteric helix. Formally, this result follows from the orthogonality condition
\[
{\bm n}\cdot{\bm\varphi}=0,
\]
which couples the cholesteric director to the vector order parameter of the $N_{TB}^*$ phase.

Finally, we note that the relationship between chirality at different, mutually incommensurate length scales established in this work may prove useful beyond the theoretical interpretation of the experimental observations reported in Ref.~\cite{AC25}. One possible example is the unusual dependence of the optical dichroism of cholesteric phases formed by short DNA fragments on osmotic pressure, as discussed in Ref.~\cite{YS17}. Another, seemingly unrelated but technologically important, example is provided by twistronic systems \cite{VA22}--\cite{OV24}, where controlling both the magnitude and the sign of chirality in layered metamaterials plays a central role.
We believe that the mechanism proposed here is generic and extends well beyond twist-bend liquid crystals to other systems exhibiting chirality on multiple, widely separated length scales

\subsection*{Acknowledgments}
I am strongly indebted to R.D.Kamien, who drew my attention to the effects of mutual influence of frustrated types of hexatic and nematic chiralities.


\end{document}